\documentclass[twocolumn,showpacs, amsmath,amssymb]{revtex4}
\usepackage{amssymb}
\usepackage[dvips]{graphicx}

\begin{document}

\title{Theory of the electron relaxation in metals excited by an ultrashort optical pump}
\author{V. V. Baranov and V.V. Kabanov}

\affiliation{Department for Complex Matter, Jozef Stefan Institute, 1001 Ljubljana, Slovenia}

\begin{abstract}
The theory of the electron relaxation in simple metals excited by an ultrashort optical pump is
developed on the basis of the solution of the linearized Boltzmann kinetic equation. The kinetic equation
includes  both the electron-electron and the electron-phonon collision integrals  and assumes that Fermi
liquid theory is applicable for the description of a simple metal. The widely used two temperature model
follows from the theory as the limiting case when the thermalization due to the electron-electron collisions
is fast with respect to the electron-phonon relaxation. It is demonstrated that the energy relaxation has two
consecutive processes. The first and most important step describes the emission of phonons by the photo-excited
electrons. It leads to the relaxation of 90\% of the energy before the electrons become thermalized among
themselves. The second step describes electron-phonon thermalization and may be described by the two
temperature model. The second stage is difficult to observe experimentally because it involves the transfer
of only a small amount of energy from electrons. Thus the theory explains why the divergence of the relaxation
time at low temperatures has never been observed experimentally.

\end{abstract}


\maketitle

\section{Introduction}
Investigations of ultrafast nonequilibrium dynamics in metals, superconductors and other strongly correlated systems after excitation by an ultrashort
laser pulse have attracted a lot of attention during the last couple of decades. The particular interest
to this field of research is related to the possibility to obtain unique information on the strength
of electron electron (e-e) and of electron-phonon (e-ph) interactions in metals and superconductors.

Up to now detailed experimental data on relaxation processes are available for metals \cite{brorson,schoenlein,elsayed,groeneveld1,brorson1},
high temperature superconductors \cite{eesley,han,chekalin,albrecht,stevens,demsar,kabanov,christoph}, and pnictide
superconductors \cite{mertelj,rettig} using standard optical pump optical probe technique. Recently a comprehensive
analysis of experimental data on optical-pump broad-band probe in high temperature superconductors has been
performed \cite{gianetti}. Most of the data are analyzed in the framework of the so-called two temperature
model (TTM) \cite{kaganov, anisimov} and e-ph coupling constants were obtained by using P.B. Allen's
theory \cite{allen}, which relates the e-ph relaxation time with the second moment,
$\lambda\langle\omega^2\rangle=2\int_0^{\infty}\alpha^2 F(\omega)\omega d\omega$,
of the Eliashberg function $\alpha^2 F(\omega)$ \cite{eliashberg2, eliashberg4}.
The  basic assumption of the model is that electrons and phonons are in a thermal quasi-equilibrium (QE)  at two
different time-dependent temperatures $T_{e}(t)$ and $T_{l}(t)$, respectively. This assumption is correct if the
e-e thermalization occurs on a much shorter timescale than the e-ph relaxation. Indeed the QE electron distribution function,
characterized by the nonequilibrium electronic temperature $T_{e}(t)$, nullifies the
electron-electron collision integral in the Boltzmann kinetic equation (BKE) and the remaining e-ph collision integral leads
to the thermalization between electrons and phonons.
However this approach has severe problems.
The e-ph thermalization is  $\tau_{e-ph}^{-1}\propto T^{3}/\omega_{D}^{2}$ in the
low $k_B T<\hbar\omega_{D}$ limit and $\tau_{e-ph}^{-1}\propto \omega_{D}^{2}/T$ in the high $k_B T >\hbar \omega_{D}$
limit\cite{allen,kabanovAlex}, where $\omega_{D}$ is the Debye frequency, and $T$ is equilibrium temperature.
As it was demonstrated in Refs.\cite{groeneveld1,kabanovAlex,Fann} the e-e thermalization rate
$\tau_{e}^{-1}\propto T^{2}/E_{F}$, where  $E_{F}$ is the Fermi energy, is much smaller than the electron-phonon
thermalization rate in the temperature range where most of the experiments are performed
$\hbar^2\omega_{D}^{2}/E_{F}<k_B T<\hbar\omega_{D}(E_{F}/\hbar\omega_{D})^{1/3}$. Therefore the main assumption of the TTM is not justified.

Experimentally it was demonstrated that the electron distribution function of a laser-heated metal is a
nonthermal distribution on the time scale of the e-ph relaxation time \cite{groeneveld1}. It was shown that
in Ag and Au the intensity dependence of the pump-probe signal as well as the temperature dependence of the relaxation
time are inconsistent with the TTM \cite{groeneveld1}. More convincing arguments against the TTM were obtained by the direct
measurements of the electron distribution function using time-resolved photoemission spectroscopy
\cite{Fann,lisowski,perfetti}. The interpretation of the photoemission spectra in terms of the
distribution function of electrons assumes that the matrix element involved in the photoemission
experiments and the density of electronic states near $E_F$ are smooth functions of energy \cite{Hufner}.
In the case of ordinary metals this assumption is usually correct. In some metals additional bands
appear in the spectra, but usually these bands have relatively large excitation energy and do not influence the
determination of the energy dependence of the distribution function (see Fig.5 in Ref.\cite{lisowski} for example).
In the measurements \cite{Fann,lisowski,perfetti} the transient electron distribution function is not thermal and has high energy tails which
survive till the thermalization occurs \cite{Fann}. Moreover in gold at about 400 fs after excitation about 30\% of the pump energy is
already in the phonon subsystem\cite{Fann}, while the thermalization is observed only after 1ps. Therefore, the transfer of energy from
electrons to phonons occurs much faster than the electronic thermalization. A similar effect is observed in
Ru \cite{lisowski}. It is estimated that 100 fs after the pump about 20\% of quasiparticles are in the high energy
tails of the distribution function. To account for this effect it was suggested that the electron
distribution function can be represented as a sum of thermal and nonthermal parts \cite{Fann,lisowski}.
It was also suggested to approximate the nonthermal distribution function by a Fermi-Dirac function
with reduced amplitude and a nonphysical auxiliary temperature \cite{lisowski}. We argue
that such a decomposition is unphysical and cannot describe the electron energy relaxation.
Instead the distribution function should be determined as a solution of the complete set of BKEs.

The accurate comparison of the TTM with experimental data on Ag and Au is presented in Ref.
\cite{groeneveld1}. It was experimentally demonstrated that the relaxation time decreases with temperature
and does not show any pump intensity dependence contrary to the TTM predictions. In order to account for
the discrepancy between the TTM and experimental data the nonthermal electron model was introduced. This model
represents the BKE for a nonequilibrium electron distribution function with electron-electron
and electron-phonon collision integrals, while assuming that phonons are at equilibrium. Numerical
integration of the BKEs allowed to reproduce experimental observations of the pump-probe
experiments on Ag and Au. Since experimental data do not demonstrate any nonlinear effects as a function
of the pump intensity the integro-differential BKE for the electrons
was reduced to differential form \cite{kabanovAlex} . The reduced equation is integrable and has an analytic solution. As a result
it was shown again that the relaxation time should increase at low temperatures \cite{kabanovAlex}, contrary to the
experimental results \cite{groeneveld1}.

Recent analysis of the pump-probe experiments on the superconducting MgB$_{2}$ \cite{jure} and
La$_{2-x}$Sr$_{x}$CuO$_{4}$ \cite{primoz} has demonstrated that the superconducting order parameter is
reduced due to the nonequilibrium phonons generated by the photoexcited carriers. It indicates that
on the sub-picosecond scale the reabsorption of the nonequilibrium phonons with the creation of
low energy electron-hole pairs is an important process which influence the energy relaxation in metals and superconductors.
Note that within the nonthermal electron model, used in Refs. \cite{groeneveld1,kabanovAlex},
this mechanism of relaxation is absent, because phonons are considered to be in equilibrium.

In this paper we develop the theory where both nonthermal electron and phonon distribution functions
are obtained by solving of linearized BKE. The e-e collisions are described
on the basis of the theory, developed in Ref.\cite{kabanovAlex}, which explicitly accounts for the conservation
of energy. In the second paragraph we derive general linearized BKE's. Then the high temperature limit of the BKEs is derived for the Eliashberg function
$\alpha^2F(\omega) \propto \omega$, which is valid for disordered simple metals, where electrons
interact with the acoustic phonons. By the direct simulations of the linearized BKEs with the different
Eliashberg functions $\alpha^2F(\omega) \propto \omega^2$ (Debye model for the acoustic phonons) and
$\alpha^2F(\omega) \propto \delta(\omega-\omega_0)$ (Einstein model for the optical phonons) we demonstrate
that the derived Fokker-Planck equations describe well the energy relaxation in the high and low temperature range.  It was also demonstrated that the energy relaxation
is not sensitive to the particular form of the Eliashberg function and is determined by the
second moment $\lambda\langle\omega^2\rangle$ of $\alpha^2F(\omega)$.
Note that recent results on the time-resolved angle-resolved photoemission spectroscopy (tr-ARPES)
indicate that the return to equilibrium of electronic excitations is determined by
the momentum and energy dependent equilibrium self-energy\cite{devereaux}.
Therefore the particular form of the Eliashberg function may be resolved in the tr-ARPES experiments. The accurate solution of
the BKE, presented in the fourth paragraph,  leads to a distribution function which is very
similar to that observed in the time-resolved photoemission
experiments\cite{Fann,lisowski,perfetti}.
Then we show that the energy transfer from the photoexcited electrons takes place on a much
shorter time scale than e-e and e-ph thermalization. Therefore the main experimentally observed
process is
determined by the emission of phonons by the photoexcited electrons and not by the e-e or e-ph
thermalization as assumed in the TTM

\section{Main equations}\label{sec2}
To derive linearized BKEs, describing both electron and phonon distribution
functions we start from the general BKE which takes both e-ph and e-e collision integrals into account. According
to Ref.\cite{stat_fiz2} the applicability of the BKE in metals is restricted by two inequalities.
1. $k_F >> 1/l$. Here $k_F$ is the Fermi momentum and $l$ is the characteristic size of inhomogeneity
of the distribution function.  In our case $l$ is restricted by the penetration depths of light or the
thickness of the film. Both of them are larger than interatomic distance and therefore this condition is
fulfilled.  2. $\hbar/\tau << E_F$. Here $\tau$ is the characteristic timescale of the changes of the
distribution function.  Therefore the application of the BKE for the description of the pump-probe experiments
in metals is justified on the timescale $t > \hbar/E_F =0.1-1$ fs, well below the time resolution
 of all known experimental data discussed here. This quasi-classical description
neglects any effects of quantum coherence which are not important in the case of simple metals. Therefore any
effects related to the quantum coherence and dephasing are out of scope of this paper.
The BKE for electrons reads:

\begin{equation}
\dot{f}_{\zeta} =I_{e-e}+I_{e-ph},
\label{kinelec}
\end{equation}
where $f_{\zeta}$ is the electron distribution function, averaged over the surface of constant energy:
\begin{equation}
f_{\zeta}= \frac{1}{N(\zeta)}\sum_{\mathbf{k}} \delta(\zeta_{\mathbf{k}}-\zeta)f_{\mathbf{k}},
\label{eldistr}
\end{equation}
Note that this averaging is justified if the distribution function depends only on the  electron energy
$f_{\mathbf k}=f_{\zeta_\mathbf k}$ and does not depend on the direction of the momentum ${\mathbf k}$.
For excitation by a spatially uniform fast optical pulse it is a reasonable assumption. If the pulse did
not penetrate the sample fully, the drift and the field terms should be included in the equation. In that
case the distribution function will be dependent on the direction of ${\mathbf k}$ and the expansion defined
by Eq.(\ref{eldistr}) is not justified. In that case more accurate expansions should be applied\cite{SmithJensen}.
Here $N(\zeta)\approx N(0)=mk_{F}/2\pi^2\hbar^2$ is the density of electronic states per spin,
$m$ is the effective mass of electron, and $\zeta$ is the electron energy counted from $E_{F}$. The density of
states is a very weak function of energy and we assume that it is constant.
The e-e collision integral has the form:
\begin{eqnarray}
&&I_{e-e}=\int\int \int d\zeta' d\epsilon d\epsilon'
K(\zeta,\zeta',\epsilon,\epsilon')\delta(\zeta+\epsilon-\zeta'-\epsilon')\times \cr
&& \left[f_{\zeta'}f_{\epsilon'}(1-f_\zeta)(1-f_\epsilon) - f_\zeta f_\epsilon (1-f_{\zeta'})(1-f_{\epsilon'})\right]
\label{ecollision2}
\end{eqnarray}
with the kernel $K(\zeta,\zeta', \epsilon, \epsilon')$ defined as:
\begin{eqnarray}
&&K(\zeta,\zeta', \epsilon, \epsilon')={2\pi\over{\hbar N(\zeta)}}\sum_{\bf k,p,q}  V_c^2({\bf q})
\delta(\zeta_{\bf k}-\zeta)\times \cr
&& \delta(\zeta_{\bf p}-\epsilon) \delta(\zeta_{\bf
k+q}-\zeta')\delta(\zeta_{\bf
p-q}-\epsilon').
\end{eqnarray}
Here $V_c({\bf q})$ is the Fourier component of the effective e-e potential.
Since we consider the relaxation of nonequilibrium electron-hole excitations with energies less than $E_{F}$
we neglect all energy dependence of the kernel $K(\zeta,\zeta^{'},\epsilon,\epsilon^{'})\approx K\approx\pi\mu_c^2/2\hbar E_F$,
where  $\mu_c$ is the Coulomb pseudopotential\cite{kabanovAlex}. The electron-phonon collision integral reads:
\begin{eqnarray}
I_{e-ph}&=&2\pi \int d\omega \int d\zeta' Q(\omega,\zeta,\zeta')\times \cr
&& \{\delta(\zeta-\zeta'-\hbar\omega)[(f_{\zeta'}-f_\zeta)\mathcal{N}_{\omega}-f_\zeta (1-f_{\zeta'})]\cr
&+& \delta(\zeta-\zeta'+\hbar\omega)[(f_{\zeta'}-f_{\zeta})\mathcal{N}_{\omega}+f_{\zeta'}(1-f_{\zeta})]\}. \label{colaverage}
\end{eqnarray}
Here $\mathcal{N}_{\omega}$ is the phonon distribution function averaged over the surface of constant frequency:
\begin{equation}
\mathcal{N}_{\omega}= \frac{1}{D(\omega)} \sum_{\mathbf{q}} \delta (\omega_\mathbf{q}-\omega) \mathcal{N}_{\mathbf{q}},
\label{phdistr}
\end{equation}
with the density of phonon states $D(\omega)=9\omega^2/\omega_{D}^3$ in the Debye approximation.
 This averaging is also possible if the excitation pulse is spatially homogeneous and
gradient terms may be omitted in the kinetic equation for phonons. The kernel of the e-ph interaction is defined as:
\begin{eqnarray}
Q(\omega,\zeta,\zeta')&=&{1\over{\hbar N(\zeta)}}\sum_{\bf k, q} M^2({\bf q})
\delta(\zeta_{\bf k-q}-\zeta')\cr
&&\delta(\zeta_{\bf k}-\zeta) \delta(\omega_{\bf
q}-\omega).
\end{eqnarray}
Here $M({\bf q})$ is the matrix element of e-ph interaction.
Because the characteristic energy of electron-hole excitations are much less
than the Fermi energy we neglect the dependence of $Q(\omega,\zeta,\zeta')$ on $\zeta,\zeta'$. As a result the e-ph
collision integral is expressed in terms of the Eliashberg function\cite{allen,kabanovAlex}:
\begin{equation}
Q(\omega,\zeta,\zeta')\approx Q(\omega,0,0)\equiv
\alpha^2 F(\omega)
\end{equation}
The kinetic equation for the phonon distribution function $\mathcal{N}_{\omega}$ reads\cite{fiz_kin}:
\begin{eqnarray}
\dot{\mathcal{N}}_\omega\!\!\!\!&=\!\!\!\!&4\pi \int_{-\infty}^{\infty} d\zeta' \int_{-\infty}^{\infty} d\zeta
Q_{ph}(\omega,\zeta,\zeta') [f_\zeta(1-f_{\zeta'})\nonumber \\
&& (1+\mathcal{N}_\omega)-f_{\zeta'}(1-f_\zeta)\mathcal{N}_\omega] \delta (\zeta'- \zeta+ \hbar\omega).
\label{kinphon}
\end{eqnarray}
 Note that here we neglect the anharmonic scattering of phonons which may be important at high temperatures.
Here we assume that the relaxation in the phonon subsystem is described by the inelastic  phonon-electron
scattering. In general the anharmonic effects may lead to an additional temperature dependence of the relaxation time at high temperatures.  The
phonon-electron kernel $Q_{ph}(\omega, \zeta, \zeta')  \approx Q_{ph}(\omega,0,0)$ is expressed in terms of the Eliashberg function:
\begin{equation}
Q_{ph}(\omega,0,0)  = Q(\omega,0,0) \frac{N(0)}{D(\omega)}=\frac{\omega_{D}^2 \beta}{2\hbar}\frac{\alpha^2 F(\omega)}{\omega^2},
\end{equation}
here
\begin{equation}
\beta=2N(0)\hbar\omega_D/9 \sim\hbar\omega_D/E_{F} \ll 1.
\label{def-beta}
\end{equation}

Since in most of the experiments the pump-probe response is a linear function of the
pump intensity \cite{groeneveld1,christoph} we linearize the kinetic equations. The electron and
phonon distribution function have the form:
\begin{equation}
f_{\zeta}=f^0_{\zeta}+\phi(\zeta,t),
\label{eldistf}
\end{equation}
\begin{equation}
\mathcal{N}_\omega=\mathcal{N}_\omega^0+\eta(\omega,t),
\label{phdistf}
\end{equation}
where $\phi(\zeta,t)$ and $\eta(\omega,t)$ are small nonequilibrium corrections to the equilibrium distribution
functions of electrons $f^0_{\zeta}=(e^{\zeta/k_B T}+1)^{-1}$ and phonons
$\mathcal{N}_\omega^{0}=(e^{\hbar\omega/k_B T}-1)^{-1}$, respectively.

In order to simplify the following calculations we introduce the dimensionless electron energy
\begin{equation}
\xi=\zeta/k_B T
\end{equation}
 and dimensionless phonon frequency
\begin{equation}
 \nu=\hbar\omega/k_B T.
\end{equation}
Therefore the functions $\phi(\zeta,t) \to \phi(\xi,t)$ and $\eta(\omega,t) \to \eta(\nu,t)$.

Let us consider first the linearized e-e collision integral. The linearized BKE was derived in Ref.\cite{kabanovAlex} and reduced
to a differential form applying Fourier transform over energy $\xi$ (see also Ref.\cite{SmithJensen},
where a similar equation is derived for the e-e collision integral). This form of the collision integral is
very useful if we consider e-e collisions only.
Consideration of the e-ph interaction in the high temperature limit leads to the differential
form of the collision integral as a function  of energy $\xi$. Therefore
it is more convenient to rewrite the e-e collision integral as a function of energy $\xi$.
Indeed, the linearized e-e collision integral has the form (Eq.(7) in Ref.\cite{kabanovAlex}):
\begin{eqnarray}
&&I_{e-e}={1\over{\pi^2\tau_e}}\Bigl [-\phi(\xi,t)(\pi^2+\xi^2)+\cr
&& 3\int_{-\infty}^{\infty} d\xi'\phi(\xi',t)\Phi(\xi,\xi')\Bigr ],
\label{ecollision3}
\end{eqnarray}
where $\tau_{e}=2/K(\pi k_B T)^2$ is the e-e thermalization time, and the kernel $\Phi(\xi,\xi')=
(\xi'-\xi)(\coth((\xi'-\xi)/2)+\tanh(\xi/2))$.
At large $|\xi-\xi'| >>1$ $\Phi(\xi,\xi')\to |\xi-\xi'|$ therefore we can simplify the e-e collision integral:
\begin{eqnarray}
&&I_{e-e}={1\over{\pi^2\tau_e}}\Bigl [-\phi(\xi,t)(\pi^2+\xi^2)+\cr
&& 3\int_{-\infty}^{\infty} d\xi'\phi(\xi',t)\Phi_1(\xi-\xi')+6\Bigl (E_1(\infty)\tanh({\xi\over{2}})-\cr &&E_1(\xi)-\xi\bigl (E_0(\infty)-E_0(\xi)\bigr )\Bigr )\Bigr ],
\label{ecollision4}
\end{eqnarray}
where the kernel $\Phi_1(x)={|x|\exp(-|x|/2)\over{\sinh(|x|/2)}}$, and $E_{\alpha}(\xi)=\int_0^{\xi}dx x^{\alpha}\phi(x,t)$. $E_{\alpha}(\xi)$
represents the dimensionless density ($\alpha=0$) and energy ($\alpha=1$) of nonequilibrium electrons with the energy less than $\xi$.
This form of the e-e collision integral Eq.(\ref{ecollision4}) is very easy to treat numerically since the integral part
represents a convolution of the distribution function $\phi$ and the kernel $\Phi_1$.

Let us turn back to the e-ph collision integrals. Substituting Eqs. (\ref{eldistf},\ref{phdistf}) to Eqs.(\ref{kinelec}, \ref{kinphon}) yields:
\begin{eqnarray}
\dot{\phi}(\xi,t)=&F_1[\phi]+F_2[\eta]; & \label{ph}\\
\quad \dot{\eta}(\nu,t)=&F_3[\phi]+F_4[\eta].& \label{th}
\end{eqnarray}
Here the expressions for $F_1$ to $F_4$ have the following form:
\begin{eqnarray}
F_1[\phi]\!\!\!\!&=\!\!\!\!&-\frac{\phi(\xi,t)}{\tau_{1}(\xi)} +\frac{2\pi k_{B}T}{\hbar \cosh(\xi/2)} \int_{-\infty}^{\infty} d\xi' sign(\xi-\xi') \nonumber \\
&&\times\alpha^2 F\Big(\frac{k_{B}T |\xi-\xi'|}{\hbar}\Big)\frac{\cosh(\xi'/2)}{2\sinh\frac{\xi-\xi'}{2}} \phi(\xi',t) \nonumber \\
&&+I_{e-e},\nonumber \\
F_2[\eta]\!\!\!\!&=\!\!\!\!&\frac{2\pi k_{B}T}{\hbar} \int_{0}^{\infty} d\nu \alpha^2 F\Big(\frac{k_B T\nu}{\hbar}\Big) \nonumber \\
&&\times\frac{\sinh^2(\nu/2)\tanh(\xi/2)}{\cosh(\frac{\xi+\nu}{2}) \cosh(\frac{\xi-\nu}{2}) }\eta(\nu,t),\nonumber\\
F_3[\phi]\!\!\!\!&=\!\!\!\!&\frac{1}{2\tau_2(\nu)\nu}
\int_{-\infty}^{\infty} d\xi' \phi(\xi',t) \frac{\sinh(\xi')}{\cosh(\frac{\nu+\xi'}{2})\cosh(\frac{\nu-\xi'}{2})}, \nonumber \\
F_4[\eta]\!\!\!\!&=\!\!\!\!&-\frac{\eta(\nu,t)}{\tau_2(\nu)}.\nonumber
\end{eqnarray}
The energy dependent electron-phonon relaxation rate
$\tau_1(\xi)^{-1}$ is defined as:
\begin{eqnarray}
&&\tau_{1}(\xi)^{-1}=\frac{2\pi k_{B}T}{\hbar} \int_{0}^{\infty} d\nu \alpha^2 F\Big(\frac{k_{B}T \nu}{\hbar}\Big)\nonumber \\
&&\times\left[\frac{1}{\sinh(\frac{\nu}{2})\cosh(\frac{\nu}{2})}+ \frac{\sinh^2(\xi/2) \tanh(\nu/2)}
{\cosh(\frac{\nu+\xi}{2}) \cosh(\frac{\nu-\xi}{2})}\right],
\end{eqnarray}
The frequency dependent relaxation rate of the phonons due to the collisions with electrons $\tau_2(\nu)^{-1}$ has the form:
\begin{equation}
\tau_2(\nu)^{-1}=\frac{2\pi\hbar\omega_D^2\beta\ }{k_{B}T} \frac{\alpha^2F(\frac{k_B T\nu}{\hbar})}{\nu}. \\
\end{equation}

Note that equation (\ref{th}) for the phonon distribution function has an analytical solution:
\begin{eqnarray}
&&\eta(\nu,t)=\frac{1}{2\tau_2(\nu)\nu} \int_{0}^{t} dt' \exp{((t'-t)/\tau_{2}(\nu))}\nonumber \\
&& \int_{-\infty}^{\infty} d\xi' \phi(\xi',t')
\frac{\sinh(\xi')}{\cosh(\frac{\nu+\xi'}{2}) \cosh(\frac{\nu-\xi'}{2})},
\end{eqnarray}
Substituting it into equation (\ref{ph}) leads to a single integro-differential equation for the distribution
function of electrons $\phi(\xi,t)$.  This equation allows only numerical analysis and is therefore not very
practical from the point of view of the analysis of experimental data. On the other hand some
assumption about the Eliashberg function and applying the high-temperature expansion leads to more simple equation.
An important simplification is obtained if we consider the model of a disordered metal with strong phonon damping.
The Eliashberg function in that limit has the following form\cite{poor1}:
\begin{equation}
\alpha^2F(\omega)=
\begin{cases}
\frac{\lambda\omega}{2\omega_D}, & \omega<\omega_D \\
0, & \omega>\omega_D
\end{cases}
\label{lowfreq}
\end{equation}
This form of the Eliashberg function leads to a frequency independent phonon-electron relaxation rate
$\tau_2^{-1}=\pi\lambda\omega_D\beta$, where the electron-phonon coupling
$\lambda$ is defined as:
\begin{equation}
\lambda=2\int_0^{\infty}d\omega\frac{\alpha^2 F(\omega)}{\omega}.
\end{equation}
The above equations are valid for any temperature $T$.
Since most of the time-resolved photoemission experiments in metals are performed at temperatures above
100K in the next section we consider the high temperature limit of the kinetic
equation Ref.\cite{kabanovAlex}.

\section{High-temperature limit}
In the high temperature limit $k_B T > \omega_D$ the BKEs (\ref{ph},\ref{th}) can be further simplified.
Let us first consider equation (\ref{th}) for the phonon distribution function.
In the high-temperature limit in the expression for $F_{3}[\phi]$ we can neglect $\nu$ under the
integral over $\xi'$. As the result the equation for $\eta(\nu,T)$ has the form:
\begin{equation}
\dot{\eta}(\nu,t)=-{\eta(\nu,t)\over{\tau_2}}+{2I(t)\over{\nu\tau_2}},
\label{ph-ht1}
\end{equation}
where the function $I(t)$ is defined by the equation:
\begin{equation}
I(t)=\int_{0}^{\infty}d\xi \tanh{(\xi/2)}\phi(\xi,t).
\label{ph-ht}
\end{equation}
This function has a very simple meaning. It describes the rate at which electrons are losing their energy to phonons (Eq.(36) in Ref.\cite{kabanovAlex}).
Equation (\ref{ph-ht1}) defines the frequency
dependence of the nonequilibrium phonon distribution function.
Indeed, if we substitute
\begin{equation}
\eta(\nu,t)=2p(t)/\nu
\label{def_eta}
\end{equation}
 to Eq.(\ref{ph-ht1}) we obtain an ordinary differential equation for the function $p(t)$:
\begin{equation}
\dot{p}(t)=-{(p(t)-I(t))\over{\tau_2}}.
\label{pott}
\end{equation}
It means that at high temperatures, where $\mathcal{N}_\omega^0\approx T/\omega$, phonons are always described
by the quasiequilibrium distribution function
and the function $p(t)$ describes the time evolution of the nonequilibrium phonon temperature.
Substituting the phonon distribution function Eq.(\ref{def_eta}) to the equation for $F_2[\eta]$
leads to the generalized Fokker-Planck equation for the nonequilibrium electron distribution function $\phi(\xi,t)$:
\begin{eqnarray}
&&\gamma^{-1} \dot{\phi}(\xi,t)=\frac{\partial}{\partial \xi}\Big[\tanh\Big(\frac{\xi}{2}\Big)\phi(\xi,t)+\frac{\partial}{\partial\xi}\phi(\xi,t)\Big]+\nonumber\\
&&\frac{p(t)\sinh(\xi/2)}{2\cosh^3(\xi/2)}
+\gamma^{-1}I_{e-e},
\label{main-ht}
\end{eqnarray}
where
\begin{equation}
\gamma=\frac{\pi\hbar\lambda \langle \omega^2 \rangle}{k_{B}T}
\end{equation}
is the e-ph relaxation rate, and the e-e collision integral is given by Eq.(\ref{ecollision4}). The
detailed derivation of the differential form for $F_1[\phi]$ is presented in Refs.\cite{kabanovAlex,gusevWright}.

 Note that equations (\ref{pott},\ref{main-ht}) are derived for the case of a disordered metal with
electrons interacting with acoustical phonons. In the Appendix we present the results of the numerical
simulations of equations (\ref{ph},\ref{th}) with three different types of Eliashberg functions at high
$T > T_D$ and low $T < T_D$ temperatures, here $T_D=\hbar\omega_D/k_B$ is the Debye temperature. We  demonstrate that equations
(\ref{pott},\ref{main-ht}) describe the relaxation of the photo-excited electrons with the accuracy
more than 10\% in the whole temperature range provided that
$\lambda \langle \omega^2 \rangle$ is constant for different Eliashberg functions. It proves that
Eqs.(\ref{pott},\ref{main-ht}) are useful because they are insensitive to the approximations and assumptions
that were used. It justifies their applicability for the description of the energy
relaxation of optically excited electrons in a large variety of ordinary metals in the whole temperature range.

Equations (\ref{pott},\ref{main-ht}) describe the relaxation of both phonon and electron distribution
functions after perturbation. These equations represent the generalization of the
TTM. The TTM may be derived from these equations in the limit when e-e relaxation is much faster than
e-ph relaxation, i.e. when $\gamma\tau_{e} \ll1$.
It is easy to check that the equilibrium distribution functions, corresponding to the simultaneous increase
of the temperature $\Delta T$ of electrons $\phi(\xi)={\Delta T\over{4T}}{\xi\over{\cosh^2{(\xi/2)}}}$ and
phonons $\eta(\nu)={\Delta T\over{T\nu}}$, represent a solution of Eqs. (\ref{pott},\ref{main-ht}).
Moreover, it is easy to check that the energies accumulated in the phonon and electron systems are proportional
to the phonon and electron specific heats. Indeed, the energy accumulated in the phonon system is
$3k_B\Delta T$. The energy in the electronic system is $2\pi^2N(0)k_B^2T\Delta T/3$. Therefore, the energies
in the phonon and electron systems are exactly proportional to their specific heats.

From Eqs.(\ref{pott},\ref{main-ht}) we can define two dimensionless parameters. The first parameter
\begin{equation}
\kappa_1=\gamma\tau_2={\hbar\lambda\langle\omega^2\rangle\over{\lambda \omega_D k_B T\beta}}\sim E_F/k_B T\gg1.
\end{equation}
The parameter $\kappa_1$ describes the relative time of generation of the electron-hole pairs at low energy by
the hot phonons which are created by electrons at large energies. This parameter is explicitly related to the electronic $C_e$
and phonon $C_{ph}$ specific heats: $\kappa_1={\pi^2C_{ph}\over{3C_e}}$. The second parameter
\begin{equation}
\kappa_2=\gamma\tau_{e}={4\hbar^2\lambda\langle\omega^2\rangle E_{F}\over{\mu_c^2\pi^2(k_B T)^3}}\gg1.
\end{equation}
It describes
the relative time of the electron thermalization due to e-e collision.
The ratio ${\kappa_2\over{\kappa1}}\sim {\hbar^2\lambda\omega_D^2\over{\mu_c^2(k_B T)^2}}$ does not contain
small parameter. The order of magnitude of these dimensionless parameters for ordinary metals follows from the
ratio of specific heats of electrons and phonons: $\kappa_1\approx\kappa_2\approx \pi^2C_{ph}/3C_e\approx 100$.
Note that parameter $\kappa_1$ is temperature dependent.

\section{Results}

In this section we discuss the results of the high temperature limit of the theory and demonstrate TTM results obtained
from the Eqs.(\ref{pott},\ref{main-ht}) in the limit of the fast e-e relaxation. We also show that the process of emission
of phonons dominates the relaxation in the low temperature limit as well.
We assume that excitation of electrons with an ultrashort laser pulse creates at $t=0$ a broad nonequilibrium distribution
of photoexcited electrons. The width of the distribution is of the order of the light frequency
 and is much larger than the Debye
frequency $\omega_D$.
The initial distribution function after the pump pulse is approximated by the formula
\begin{equation}
\phi(\xi,0)={\xi\over{\Omega^3}}\exp(-\xi^2/\Omega^2).
\end{equation}
Here $\Omega$ the dimensionless excitation frequency which is defined as frequency of light measured in units of $k_B T/\hbar$.
This formula preserves the energy per pulse for different frequencies of excitation $\Omega$.

 Note that this formula is different from the standard distribution of quasiparticles created by an optical
pulse\cite{elesin}. It has a characteristic energy which is of the order of the frequency of excitation. This
function allows the existence of quasiparticles
with the energy higher than $\Omega$, but it is decreasing very quick at $\xi >\Omega$. In the Appendix
we present the results of numerical simulations assuming
different initial distribution functions.  The simulations show that irrespective of the particular choice of $\phi(\xi,t=0)$,
the distribution function becomes independent on initial conditions on the time scale $t\sim 10^{-3} \tau_e$.
Therefore the particular choice of $\phi(\xi,t=0)$ does not influence the process of energy relaxation of photo-excited electrons
at the time scale longer than $10^{-3}\tau_e$.

In Fig.1 the time evolution of the electron distribution function is presented in the absence of the e-ph interaction.
The evolution is characterized by the fast reduction of the high energy part of the distribution function and fast
increase of the distribution function at small energies $\xi \sim 1$. As it follows from Fig.1 the high
energy tails of the distribution functions disappear at the time scale determined by $\tau_{e}$. This is consistent
with the previous analysis of the e-e relaxation \cite{kabanovAlex}. The time dependence of the density of nonequilibrium
electrons, presented in Fig.1 (inset), shows that electron thermalization occurs on the time scale
of $\tau_{e}$ irrespective of the photon energy $\Omega$. The difference is important only on a very short time scale
for relatively low frequency $\Omega$ and is due to the fact that the initial density of nonequilibrium electrons depends on the pump
frequency $n(0)\propto\Omega^{-1}$, provided that the energy per pulse is constant.
Here 
\begin{equation}
n(t)=\int_{0}^{\infty}d\xi \phi(\xi,t)
\label{density}
\end{equation}
is the dimensionless density of nonequilibrium electrons. The dimensionless density $n(t)$ is actually equal to the measured density
of nonequilibrium electrons expressed in units of $N(0)k_B T$.
Note that the scaling
arguments presented in Ref.\cite{tas} predict $n(t) \propto \sqrt{t}$. Our calculations do not support this behaviour.
However with the increase of $\Omega\to \infty$ the dependence of $n(t)$ becomes more and more similar to $\sqrt{t}$.
\begin{figure}
\includegraphics[width = 87mm, angle=-0]{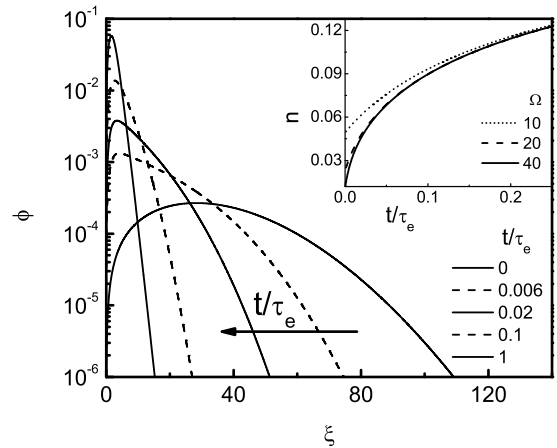}
\caption{Evolution of the nonequilibrium distribution function by e-e collisions.
The distribution function is an odd function with respect to $E_F$ ($\xi=0$). Inset demonstrates the time dependence of the
nonequilibrium electron density for three different photon energies $\Omega$.}
\end{figure}

In order to demonstrate the relation of the  TTM to Eqs.(\ref{pott},\ref{main-ht}) we consider the unphysical limit
$\kappa_2=\gamma\tau_{e}<<1$. This limit corresponds to the case of the fast electron thermalization with respect to
e-ph relaxation. The time evolution of the nonequilibrium distribution function $\phi(\xi,t)$ in this limit is plotted
in Fig.2. As it follows from Fig.2 at the time scale $t\sim \tau_e =\kappa_2\gamma^{-1}<<\gamma^{-1}$ the high energy
part of the nonequilibrium distribution function disappears and $\phi(\xi,t)={\Delta T_e(t)\over{4T}}{\xi\over{\cosh(\xi/2)^2}}$.
Here $\Delta T_e(t)$ is the change of the electron temperature. This function nullifies the e-e collision integral and
further evolution of the distribution function $\phi$ is described by the time dependence of $\Delta T_e(t)$ as it follows from Fig.2.
The $\xi$ dependence of the distribution function at $t>\tau_e$ is always the same $\phi(\xi,t) \propto {\xi\over{\cosh(\xi/2)^2}}$
and the evolution with  time is described by the time dependent pre-factor. Note that theory is linear in the excitation intensity
and therefore $\Delta T_e/T <1$, therefore the time evolution of $\Delta T_e(t)$ is described by the
linearized TTM. Note that for the case, presented in Fig.2 $\kappa_1=20\gg1,\kappa_2=0.75$ the dimensionless energy density $\Delta E\approx 0.07$ transferred
from electrons to phonons during the thermalization time $t\sim \tau_e$ is much smaller than the energy absorbed by electrons
$\Delta E \ll E(t=0)=0.443$. The dimensionless energy accumulated by nonequilibrium electrons is defined as:
\begin{equation}
E(t)=\int_{0}^{\infty}d\xi \xi\phi(\xi,t).
\label{def_energy}
\end{equation}
Therefore $\Delta T_e$ at $t\approx\tau_e$ can be evaluated using the electronic specific heat
$C_{e}\Delta T_e=E_e$. Then thermalization between electrons and phonons occurs on the time scale  $t\sim \pi^2/3\gamma$ in accordance with the TTM.
The distribution function at $t\to \infty$ corresponds to $\phi(\xi,t)={\Delta T_{\infty}\over{4T}}{\xi\over{\cosh(\xi/2)^2}}$,
where $\Delta T_{\infty}$ is found from the equation: $\Delta T_{\infty}(C_e+C_{ph})=E_e$.
Note that the measure for electronic temperature in that case is not the width of the distribution function but the value
of $\phi(\xi,t)$ at the maximum which is approximately equal to $\Delta T/4T$. The most important property of this limit is the absence of
the high-energy tails of the distribution function, which directly follows from the fact that $\kappa_2 <1$.
\begin{figure}
\includegraphics[width = 87mm, angle=-0]{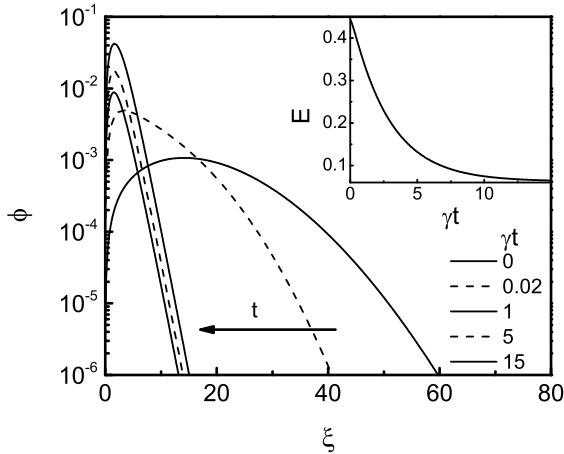}
\caption{Time dependence of the distribution function in the TTM limit $\kappa_1=20$, $\kappa_2=0.75$
and $\gamma$ is the e-ph relaxation rate. Inset shows time dependence of the electron energy.}
\end{figure}

Now let us consider the more realistic case $\kappa_1=\kappa_2=100 \gg 1$. These parameters roughly correspond to the case of metals like
Au\cite{Fann} or Ru\cite{lisowski}. The time evolution of the distribution function is presented in Fig.3.
At the timescale of $t\sim 15\gamma^{-1} \ll\tau_e$ the high
energy tail of the distribution function disappears and the electron distribution function can be approximated
by the quasiequilibrium distribution function characterized
by the electronic temperature $\phi(\xi,t)={\Delta T_{e}\over{4T}}{\xi\over{\cosh(\xi/2)^2}}$. The electronic
temperature $\Delta T_{e}$  is not defined by the conservation of energy, because by the time of "thermalization"
about 90\% of the energy has gone to phonons (See inset of Fig.3). The dimensionless energy accumulated by nonequilibrium electrons is defined in Eq.(\ref{def_energy}).
This is consistent with the time resolved photoemission data on Au\cite{Fann} and Ru\cite{lisowski}. According to
Ref.\cite{lisowski} the estimated electronic temperature at the peak $\Delta T_{e}\approx 125K$ is much less than
electronic temperature estimated from TTM $\Delta T_{e}\approx 1200K$.
The final stage of the relaxation in that case can be characterized by the slight decrease of the electronic temperature
to its equilibrium value and can be described by the TTM. Note, however that the final
stage of relaxation is very difficult to observe experimentally, because it involves the transfer of a very small amount of
energy (less than 10\% of the pump energy) from electrons to phonons. The biggest changes in the nonequilibrium
distribution function take place during the initial stage of relaxation where the TTM is not applicable.
\begin{figure}
\includegraphics[width = 87mm, angle=-0]{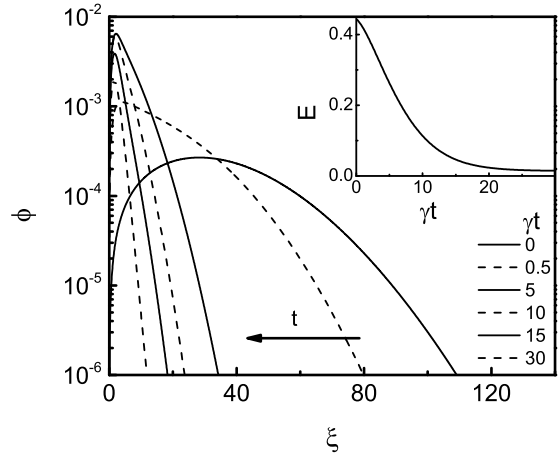}
\caption{Time evolution of the distribution function in the case of ametal with $\kappa_1=\kappa_2=100$.
Inset shows time dependence of the electron energy.
By the time of the electron thermalization 90\% of energy is in the phonon subsystem.
$\gamma$ is the e-ph relaxation rate and $\tau_e=100\gamma^{-1}$.}
\end{figure}

From these calculations the following qualitative picture of the relaxation of the photoexcited electrons emerges.
The pump pulse creates a broad distribution of electron-hole pairs with large excitation energy.
The high energy electrons relax to the low energy scale $\zeta \sim (\hbar\omega_D E_F)^{1/2}\gg k_BT,\hbar\omega_D$
due to e-e collisions. It happens on the time scale $\langle\omega\rangle/\lambda\langle\omega^2\rangle$. The photoexited
electron-hole pairs emit phonons immediately after excitation. The emission rate is temperature independent and is
not affected by the Fermi distribution function, because the average energy of nonequilibrium electrons is
large in comparison with the phonon frequency and therefore the factor $(1-f_{\xi-\nu})$ in the emission
probability can be replaced by 1.
 In the Appendix it is shown that the Fokker-Planck equation describes well energy relaxation
in the low temperature limit as well. (see Fig. 8, 9 and 10 in the Appendix). The absence of the divergence of the
relaxation time at low temperatures $k_B T< \hbar\omega_D$ for different e-e collision times is presented in Fig. 4.
The relaxation time, defined as the time when half of the energy is transferred from electrons to phonons, is weakly
temperature dependent at high temperature and temperature independent at low temperatures. Note that electron-electron
collisions have an important effect on energy relaxation. When the electron-electron collision rate increases,
the high-energy excitations are decaying with the creation of more low energy electron-hole pairs. The energy relaxation
on the other hand is proportional to the number of nonequilibrium electrons. Therefore when the e-e collision rate
increases the energy relaxation rate of the nonequilibrium electrons increases as well. This is demonstrated in Fig.4.
\begin{figure}
\includegraphics[width = 87mm, angle=-0]{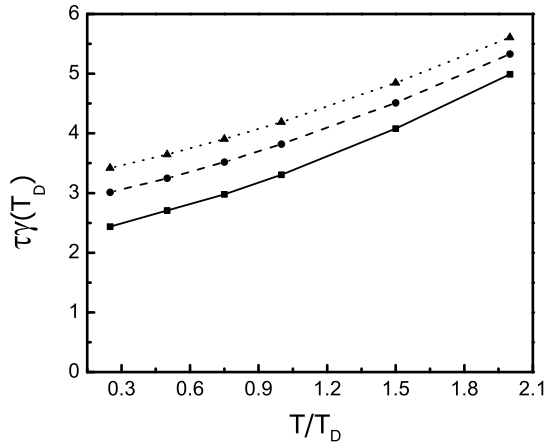}
\caption{Temperature dependence of the energy relaxation time for large $\kappa_1\to \infty$ for different
$\gamma(T_D)\tau_e(T_D)$=40  (solid line), 80 (dashed line), and 120 (dotted line).
$T_{D}=\hbar\omega_D/k_B$ is the Debye temperature.}
\end{figure}
When the average energy of nonequilibrium electrons, because of emission of phonons, is reduced to the scale
$\zeta\sim k_B T,\hbar\omega_D$ the $(1-f_{\xi-\nu})$ factor becomes important leading to the strong slowing down
of the relaxation, indicating the second stage of relaxation.
Since most of the energy is transferred to the phonon subsystem before the time when the width of the nonequilibrium
distribution function becomes of the order of $k_B T$, the final and
the longest stage of thermalization is difficult to observe.
In order to demonstrate that we multiply Eq.(\ref{main-ht}) by $\xi$ and integrate over $\xi$ from $0$ to $\infty$ with the following result:
\begin{equation}
\dot{E}(t)=-\gamma (I(t)-p(t))
\label{el-ener}
\end{equation}
$E(t)$ is defined in Eq.(\ref{def_energy}). The e-e collision integral conserves electron energy and therefore does not contribute to Eq.(\ref{el-ener}).
The first term in this equation describes the loss of the energy due to generation of phonons. The second term describes the
reabsorption of phonons with the generation of low energy electron-hole pairs and slows down the relaxation. If $\kappa_1>>1$
we can neglect the second term in this equation. When the distribution
function $\phi(\xi,t)$ is broad and has tails at $\xi >>1$, $\tanh(\xi/2)$ in the integral Eq.(\ref{ph-ht}) can be replaced to 1
and therefore $I(t)\approx n(t)$, where $n(t)$ is defined in Eq.(\ref{density}).
Therefore Eq.(\ref{el-ener})
indicates that energy loss by electrons is proportional to the nonequilibrium electron density.
In order to illustrate these arguments we plot in Fig.5  the time dependence of the electron energy for the case of
large $\kappa_1$ in comparison with the approximate formula:
\begin{equation}
E(t)=E(0)-\gamma\int_0^t dt' n(t').
\label{approx_E}
\end{equation}
As it follows from Fig.5, this approximation describes
well the time evolution of the energy until $t=(10-15)\gamma^{-1}$
when the high energy tails in the nonequilibrium distribution function disappear.
\begin{figure}
\includegraphics[width = 87mm, angle=-0]{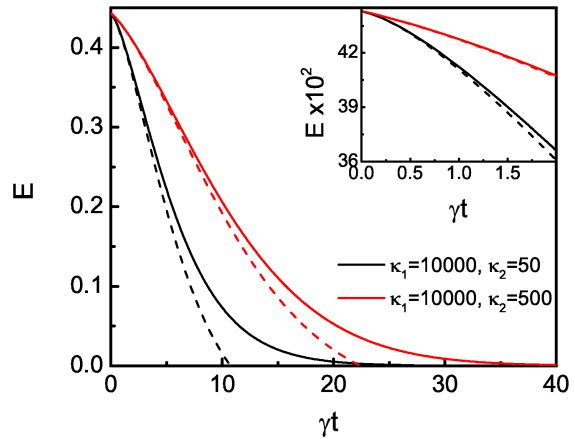}
\caption{Energy relaxation in metals with large $\kappa_1=10000$ and different e-e relaxation times given by
$\tau_e=\kappa_2\gamma^{-1}$, $\gamma$ is the e-ph relaxation rate. Dashed lines represent the approximate formula Eq.(\ref{approx_E}) (see the text).}
\end{figure}
Note that the slope of the energy relaxation curve at $t=0$ is the same in both cases (see inset of Fig.5) because the number of photo-excited
electrons is the same at $t=0$. The difference between the two cases is due to the e-e collisions. Since for the case of
$\kappa_2=50$ the quasiparticle multiplication is much faster the energy relaxation rate is larger. This is demonstrated in Fig.6.
The characteristic maximum in the quasiparticle density for the case of $\kappa_2=50$ is approximately 2 times larger than for the case
of $\kappa_2=500$. Therefore the relaxation rate is determined by the nonequilibrium quasiparticle density at relatively high energy
$\xi >1$. The quasiparticle density strongly depends on $\tau_e$. Therefore analysis of the energy relaxation after the ultrashort
pump pulse allows one to obtain not only information about the e-ph interaction constant, but also about $\tau_e$ and as a
consequence about the Coulomb pseudopotential $\mu_c$.
\begin{figure}
\includegraphics[width = 87mm, angle=-0]{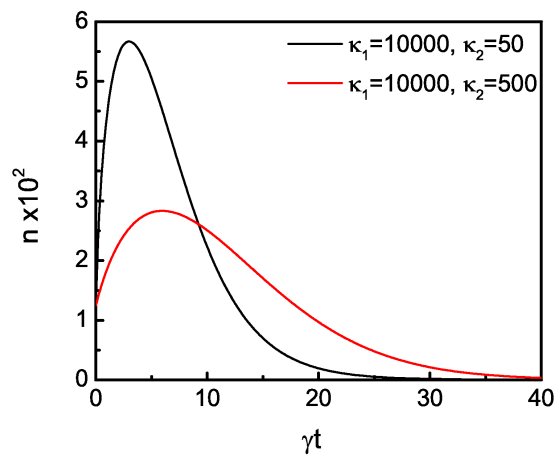}
\caption{Time evolution of nonequilibrium density for the same case as Fig.5.}
\end{figure}

The physical meaning of Eq.(\ref{el-ener}) is very simple. Every non-equilibrium electron emits phonons with the temperature independent rate
$\tau_{em}^{-1}=\pi \lambda \langle \omega^2 \rangle/\langle \omega \rangle$. The formula for the emission rate is valid in the low temperature
limit as well. Indeed, if we consider the zero temperature limit $T=0$ in Eq.(\ref{colaverage}) and calculate the phonon emission rate we obtain
the same expression for $\tau_{em}$ as in high temperature limit (for details of calculations see Ref.\cite{beyer}). Since the emission rate
for non-equilibrium electrons is much larger than the thermalization rate in both low temperature and high temperature limits, we conclude that
low temperature divergence of the relaxation time will not be observed experimentally. Most of the energy will be transferred to the phonon
subsystem before low temperature relaxation processes become important.  This is confirmed by the comparison of the results of
calculations of the temperature dependent energy relaxation time using the Focker-Planck equation (\ref{pott},\ref{main-ht})
with the results obtained from the kinetic equations (\ref{ph},\ref{th}) with different Eliashberg functions (Fig.10).

Note that the theory predicts the dependence of the relaxation rate on the pump frequency $\Omega$ (Fig.7). Indeed, preservation of
the energy per pulse leads to the increase of the number of photoexcited electrons $n\propto \Omega^{-1}$. If the pump
frequency is large enough, e-e collisions lead to the quick disappearance of quasiparticles with the energy higher than the threshold energy.
The threshold energy is defined as the energy where the life-time of a nonequilibrium electron
due to e-e collisions becomes comparable with the life-time due to e-ph collisions. Therefore excitation with the light frequency
higher than this threshold energy does not lead to the pump frequency dependence. The situation is different when the pump frequency
is below the threshold energy. In that case the e-e collisions are not important and the relaxation rate is governed by the
number of photoexcited electrons $n\propto\Omega^{-1}$, leading to the increase of the relaxation rate. This behaviour is demonstrated in Fig.7.
\begin{figure}
\includegraphics[width = 87mm, angle=-0]{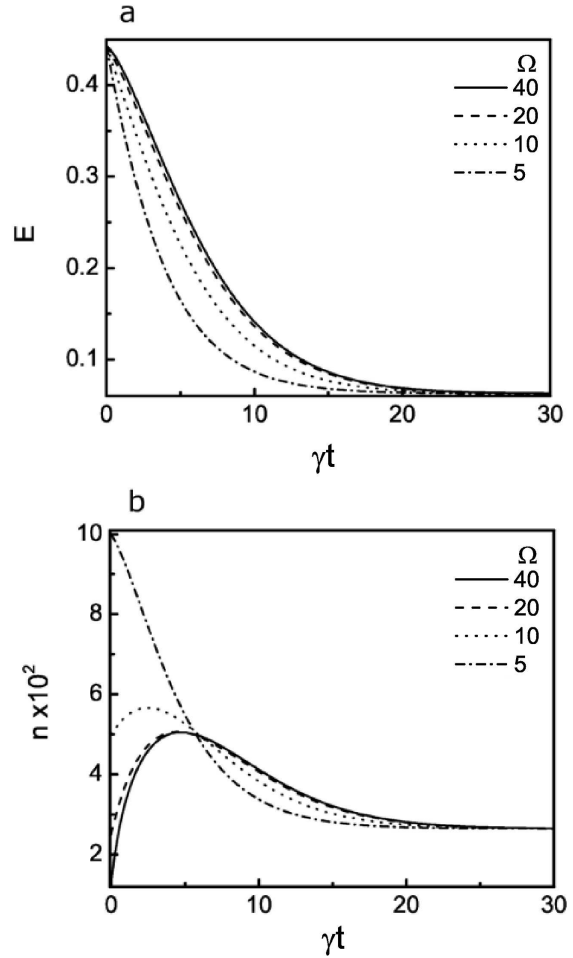}
\caption{a)Energy relaxation b) and time dependence of the nonequilibrium density for different pump frequencies $\Omega$. When the pump frequency
is relatively small the relaxation time depends on the pump frequency.}
\end{figure}

Experimental measurements of time-resolved photoemission in metals\cite{Fann,lisowski} provide direct information about
the time dependent nonequilibrium distribution function $\phi(\xi,t)$ (see Fig.7 in \cite{lisowski}). This can be directly
compared with the results of our calculations. One of the most striking resemblance between our theory and the experiments
is the existence of the high energy tails in the nonequilibrium distribution function when a substantial amount of the energy is
already transferred to the phonon subsystem (Fig.3). Very often in the experimental analysis these high energy tails are
interpreted as the nonequilibrium electron temperature $\Delta T_e$. Usually the results of the measurements of the nonequilibrium
distribution function $f_{\xi}(t)=f^0_{\xi}+\phi(\xi,t)$ are plotted on a logarithmic scale as a function of the energy $\xi$ and
the slope of $f_{\xi}(t)$ defines the nonequilibrium electron temperature $\Delta T_e(t)$ \cite{lisowski,perfetti}. In this
analysis a large part of the nonthermal electrons is accounted for thermal. We suggest a different way of defining
the electron temperature. In Fig.7 of Ref.\cite{lisowski} the nonequilibrium part of the electron distribution function
$\phi(\xi,t)$ is plotted for different time delays after pump.
From this graph it is clear that at any time delay $\phi(\xi,t) <<1$ and therefore the condition for the linearization of
the BKEs is fulfilled. If $\phi(\xi,t)$ is described by the nonequilibrium temperature then
$\phi(\xi,t)={\Delta T_e(t)\over{4T}}{\xi\over{\cosh(\xi/2)^2}}$. The shape of $\phi$ in Fig.7 of Ref.\cite{lisowski}
is similar to that described by this formula. The function $\xi/\cosh^2(\xi/2)$ has a maximum at $\xi \approx 1.5-1.6$ which
is about 1. Therefore the electronic temperature can be defined directly from the maximum of $\phi$ in Fig.7\cite{lisowski}
$\Delta T_e(t)=4T\phi_{max}$ where $\phi_{max}$ is the experimental value of the maximum of $\phi(\xi,t)$. Therefore after 100fs
$\Delta T_e\approx 28K$ contrary to the estimate from the slope of the distribution function $f(\xi,t)$ $\Delta T_e\approx 125K$.

Another important consequence of the time-resolved photoemission experiments is the possibility to evaluate the electron energy
and the number of the photoexcited electrons and holes. According to our theoretical results the maximum in energy and
maximum in the density of photoexcited electrons should be shifted in time with respect to each other (Figs.5,6).
If the pump pulse is much shorter than $\gamma^{-1}$ the energy has its maximum immediately after pump pulse, because the
photo-excited electrons cannot emit any phonon during the short pulse. The high energy photoexcited electrons reduce
the energy due to e-e collisions leading to an increase in the number of nonequilibrium electrons. The electron-hole
recombination is slow at the short time scale because the number of nonequilibrium electrons within the energy interval $\xi < \omega_D$ is small
when the pump frequency is large $\Omega >>\omega_D$. It means that the number of photoexcited electrons increases
immediately after the pump. The maximum in the nonequilibrium electron density occurs when the process of electron
multiplication due to e-e collisions is compensated by the electron-hole recombination with the emission of phonons.
The time when the density of the nonequilibrium
electrons has its maximum indicates the end of the first stage of electron thermalization. After that the energy
relaxation rate decreases. The difference in the positions of
maxima can be clearly seen in Figs. 8 and 9 of Ref.\cite{lisowski}.

Note that Eq.(\ref{el-ener}) allows the evaluation of the electron-phonon coupling constant. If we evaluate
the decay rate of the electron energy experimentally and divide it with $n(t)$ on a short time scale after
the maximum of the $E(t)$ curve this ratio
should be constant and is exactly equal to $\gamma$. Note that the determination of the coupling constant
from the single color pump-probe
measurements only is problematic. The energy relaxation rate depends on the e-e and e-ph coupling constants
as demonstrated in
Fig.4 and 5 and requires the knowledge of both $E(t)$ and $n(t)$ dependences. Therefore, independent
measurements of the nonequilibrium density are necessary to evaluate the electron-phonon
coupling constant $\lambda$.

\section{Conclusion}

We developed the theory of the electron relaxation in metals excited by an ultrashort optical pump.
The theory is based on the solution of the linearized BKE which includes  the electron-electron and the
electron-phonon collision integrals. The well known two temperature model represents the limiting case of
the theory when thermalization due to the electron-electron collisions is fast with respect to
the electron-phonon relaxation.

We demonstrated that for realistic parameters the energy transfer from electrons to phonons occurs
on a timescale which is much faster than the electron thermalization. The high energy tails in the
electron distribution function disappear when most of the energy is already transferred to phonons as
it is observed in the time-resolved photoemission experiments \cite{Fann,lisowski}. The reabsorption
of nonequilibrium phonons slows down the relaxation.

We demonstrate that the relaxation of the photo-excited electrons occurs in two steps. The first and
the most important step represents the emission of phonons by the nonequilibrium electrons. The rate of electron
energy loss at that stage is proportional to the density of nonequilibrium electrons. The temperature dependence
of the relaxation at this stage is significantly different from the predictions of the two temperature model.
The density of nonequilibrium electrons is strongly influenced by the electron-electron collisions. It makes
the relaxation time dependent on the pump frequency if the pump frequency is smaller than some threshold
frequency. It also allows the evaluation of the electron-phonon coupling constant $\lambda$ from time-resolved
photoemission data.

The second stage of the relaxation describes the electron-phonon thermalization. This stage may be described
approximately by the two temperature model. Since it involves a small energy transfer (about 10\% or less)
from electrons to phonons it is difficult to observe experimentally. Our theory explains one of the most
severe problems of the theory why the divergence of the relaxation time at low temperatures was never observed experimentally.

\section{Acknowledgments}
We are grateful to  Jure Demsar, Dragan Mihailovic,
Christoph Gadermaier, Tomaz Mertelj, Alexander Balanov,
Laurenz Rettig and Uwe Bovensiepen for sharing with us their insight into
non-equilibrium phenomena in metals and superconductors.

\section{Appendix}

In order to demonstrate that Eqs. (\ref{pott},\ref{main-ht}) are independent of the approximations that were made,
we perform extensive
numerical simulations of equations (\ref{ph},\ref{th}) in the limit of large $\kappa_1\gg 1$ with three
different Eliashberg functions: the linear function of phonon frequency $\alpha^2F(\omega)\propto \omega$ Eq.(\ref{lowfreq}), the Debye model
\begin{equation}
\alpha^2F(\omega)=
\begin{cases}
\frac{\lambda\omega^2}{\omega_D^2}, & \omega<\omega_D \\
0, & \omega>\omega_D
\end{cases}
\label{debye}
\end{equation}
and the Einstein model
\begin{equation}
\alpha^2F(\omega)=\lambda\omega_0\delta(\omega-\omega_0).
\label{einstein}
\end{equation}
Since the e-ph collision integral for all these cases is similar to the e-e collision
integral it is very easy to treat numerically. The evolution of the nonequilibrium distribution
function at low temperature $k_B T = \hbar\omega_D/4$ is presented in Fig.8.
\begin{figure}
\includegraphics[width = 87mm, angle=-0]{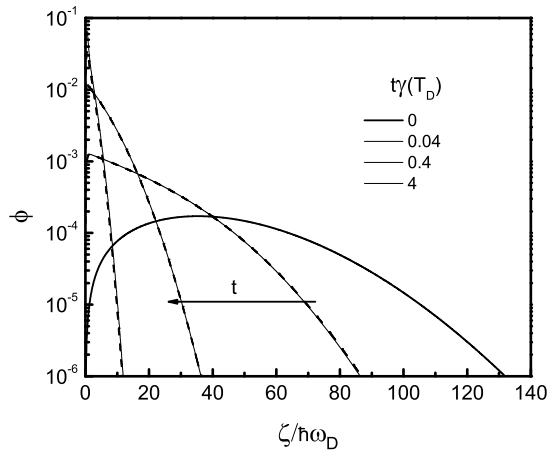}
\caption{Time evolution of the electron distribution function. Dashed line is from the solution of the Fokker-Plank equation Eq.(\ref{main-ht}) and
solid line represents the solution of the linearized
BKE Eqs.(\ref{ph},\ref{th}) with the Eliashberg function defined by Eq.(\ref{debye}). $\gamma(T_D)\tau_e(T_D)=40$.}
\end{figure}
It is easy to see that in the whole range of time the electron distribution function calculated with
Eqs.(\ref{ph},\ref{th}) with the Eliashberg function (\ref{lowfreq}) is almost undistinguishable from
the distribution function obtained by the solution of the Fokker-Planck equation (\ref{pott},\ref{main-ht}).
In Fig.9 we present the electron energy as a function of time calculated on the basis of Eqs.(\ref{ph},\ref{th})
for different $\alpha^2F(\omega)$ functions defined by Eqs.(\ref{lowfreq},\ref{debye}, and \ref{einstein})
in comparison with the solution of the Fokker-Planck equation. Here we keep $\lambda\langle \omega^2\rangle$
constant for different Eliashberg functions. Again the energy relaxation is almost the same for different cases
in the high temperature $k_B T > \hbar\omega_D$ region and differs not more than by 10\% at low temperatures
$k_B T < \hbar\omega_D$. Note that the difference between different models increases when almost all energy is transferred
to phonons and electron thermalization takes place. It is because the Fokker-Planck equation overestimates the
thermalization rate in the low temperature range. On the other hand the temperature dependence of the thermalization
rate for different types of Eliashberg functions is different. In the case of the Eliashberg function defined by Eq.(\ref{lowfreq})
the thermalization time is proportional to $\tau\propto T^{-2}$, in the Debye model $\tau\propto T^{-3}$ and in the
case of Einstein model related to the e-ph interaction with optical phonons it is an exponential function of
temperature $\tau\propto \exp(-\hbar\omega_0/k_BT)$.
\begin{figure}
\includegraphics[width = 87mm, angle=-0]{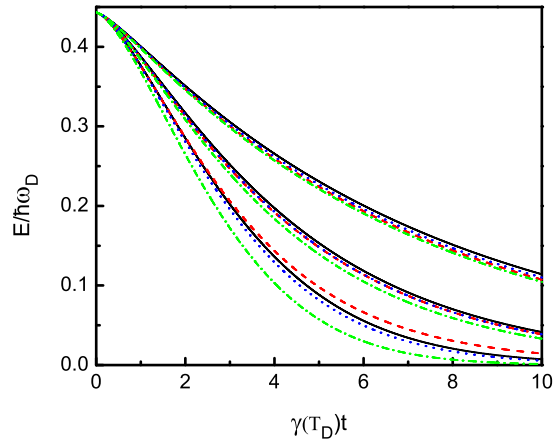}
\caption{Electron energy as function of time, calculated for different Eliashberg functions and at different temperatures $T=2T_D$ (upper series of curves), $T=T_D$ (middle series of curves), ant $T=0.3T_D$ (lower series of curves).
Solid line represents the Debye model Eq.(\ref{debye}), dashed line represents the Einstein model Eq.(\ref{einstein}), dotted line
represents the "dirty metal case" Eq.(\ref{lowfreq}) and dashed-dotted line represents the energy calculated using
Fokker-Planck equation Eq.(\ref{main-ht}).$\gamma(T_D)\tau_e(T_D)=40$. In the case of the Einstein model $T_D=\hbar\omega_0/k_B$.}
\end{figure}
Therefore we can conclude that energy relaxation of the photo-excited electrons is well described by the
Fokker-Planck equation (\ref{main-ht}) in the whole temperature region.
The relaxation does not depend on the particular form of the Eliashberg function and is defined by
the second moment of the Eliashberg function $\lambda\langle \omega^2\rangle$.
The energy relaxation is not exponential, therefore we define the relaxation time as the time at which half of the absorbed energy
is transferred from electrons to phonons.

The temperature dependence of the energy relaxation time is plotted in Fig.10 for three different Eliashberg functions
and for the Fokker-Plank equation. As it is clearly seen from this graph, there is marginal difference between these cases at temperatures $T>0.2T_D$.
Therefore the relaxation time is independent of the particular form of the Eliashberg function.
It is determined by the second moment of the Eliashberg function $\lambda\langle \omega^2 \rangle$ in this temperature range.  It is almost independent
of the particular form of the $\alpha^2F(\omega)$ function and is not sensitive to whether the acoustic or optical phonons dominate the
the relaxation, provided that $\lambda\langle\omega^2\rangle$ is constant. Note, that in the case of tr-ARPES experiments, where
the the relaxation time may be momentum dependent some of the features of the Eliashberg function may be resolved\cite{devereaux}.

In the low temperature region there is a very small
increase of the relaxation time for the Debye (2\% at $T=T_D/20$) and for the Einstein(5\% at $T=T_D/20$) models for the Eliashberg function. As it was mentioned in the introduction the two temperature model becomes valid at very low temperatures.  Since the e-ph relaxation time increases exponentially with decreasing of temperature $\tau_{e-ph}\propto \exp(\hbar\omega_0/k_BT)$ in the case of the interaction with optical phonons, the range of temperatures where the two-temperature model is efficient may be relatively broad $T<T_D/10$. In the case of interaction with the acoustic phonons this range is more narrow. Therefore if the relaxation is dominated by the interaction with the optical phonons it may be possible to observe the increase of the relaxation time at low temperatures. From an experimental point of view this is very unlikely because the interaction with the acoustic phonons will dominate the relaxation at low temperatures.
\begin{figure}
\includegraphics[width = 87mm, angle=-0]{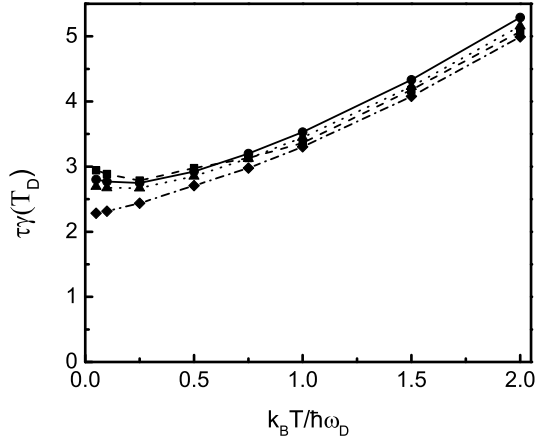}
\caption{Temperature dependence of the energy relaxation time calculated for large $\kappa_1\to \infty$ for three different Eliashberg functions.
Dotted line with triangles represents the "dirty metal case" Eq.(\ref{lowfreq}), solid line with circles represents Debye model Eq.(\ref{debye}),
and dashed line with squares represents the Einstein model Eq.(\ref{einstein}). Solid lines with diamonds represents temperature dependence of the
relaxation time for the high temperature Fokker-Planck equation Eq.(\ref{main-ht}). $\gamma(T_D)\tau_e(T_D)=40$. In the case of Einstein model $T_D=\hbar\omega_0/k_B$.}
\end{figure}

In order to show that the results are independent of the initial distribution function of
the photo-excited electrons we present the results of the time evolution of the distribution function
for different initial conditions$\phi(\xi,0)$. In Fig.11 we plot the time evolution of the distribution function
for $\phi(\xi,0)={\xi\over{\Omega^3}}\exp(-\xi^2/\Omega^2)$ compared with the case where $\phi(\xi,0)=12(\Omega-\xi)\xi/\Omega^4$.
The results clearly demonstrate that the essential difference between these two cases survives only on the time scale less than
$10^{-3}\tau_ee$ when the transfer of energy from nonequilibrium electrons to phonons is negligible. Therefore the particular
choice of the initial distribution function is not important. The only important parameter is the characteristic energy scale of
this distribution, which in both cases is described by the characteristic frequency $\Omega$ which is related to the pump frequency.
\begin{figure}
\includegraphics[width = 87mm, angle=-0]{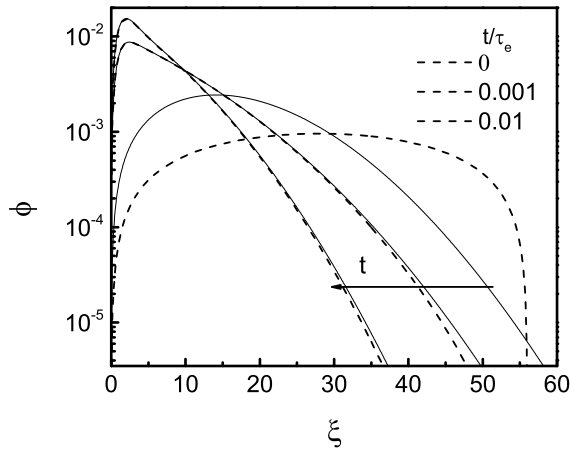}
\caption{Time evolution of the nonequilibrium distribution function for different initial
distributions of the photoexcited electrons. Solid line corresponds to exponential distribution of the photoexcited
electrons and dashed line corresponds to $\phi(\xi,0)=12(\Omega-\xi)\xi/\Omega^4$. }
\end{figure}


\begin{thebibliography}{200}
 \bibitem{brorson} S. D. Brorson, A. Kazeroonian, J. S. Moodera, D. W.
Face, T. K. Cheng, E. P. Ippen, M. S. Dresselhaus, and G. Dresselhaus,
Phys. Rev. Lett. {\bf 64}, 2172 (1990).

\bibitem{schoenlein} R. W. Schoenlein, W. Z. Lin, J. G. Fujimoto, and G. L.
Eesley, Phys. Rev. Lett. {\bf 58}, 1680 (1987).
\bibitem{elsayed} H. E. Elsayed-Ali, T. B. Norris, M. A. Pessot, and
G. A. Mourou, Phys. Rev. Lett. {\bf 58}, 1212 (1987).
\bibitem{groeneveld1} R. H. M. Groeneveld,  R. Sprik, and Ad Lagendijk, Phys. Rev. B
{\bf 51}, 11433 (1995).
\bibitem{brorson1} S. D. Brorson, J. G. Fujimoto, and E. P. Ippen, Phys.
Rev. Lett. {\bf 59}, 1962 (1987).
\bibitem{eesley} G. L. Eesley, J. Heremans, M. S. Meyer, G. L. Doll,
and S.H. Liou, Phys. Rev. Lett. {\bf 65}, 3445, (1990).
\bibitem{han} S. G. Han, Z. V. Vardeny, K. S. Wong, O. G. Symko, and G.
Koren, Phys. Rev. Lett. {\bf 65}, 2708 (1990).
\bibitem{chekalin} S. V. Chekalin, V. M. Farztdinov, V. V. Golovlyov,
V. S. Letokhov, Yu. E. Lozovik, Yu. A. Matveets, and A. G. Stepanov
Phys. Rev. Lett. {\bf 67}, 3860 (1991).

\bibitem{albrecht} W. Albrecht, Th. Kruse, and H. Kurz, Phys. Rev.
Lett. {\bf 69}, 1451 (1992).
\bibitem{stevens} C. J. Stevens, D. Smith, C. Chen, J. F. Ryan, B.
Podobnik, D. Mihailovic, G. A. Wagner, and J. E. Evetts, Phys. Rev.
Lett. {\bf 78}, 2212 (1997).
\bibitem{demsar} J. Demsar, B. Podobnik, V. V. Kabanov, T. Wolf, and
D. Mihailovic, Phys. Rev. Lett. {\bf 82}, 4918 (1999).
\bibitem{kabanov} V. V. Kabanov, J. Demsar, B. Podobnik, and D.
Mihailovic, Phys. Rev. B {\bf 59}, 1497 (1999).
\bibitem{christoph} C. Gadermaier, A.S. Alexandrov, V.V. Kabanov, P. Kusar,
T. Mertelj, X. Yao, C. Manzoni, D. Brida, G. Cerullo, and D. Mihailovic
Phys. Rev. Lett. {\bf 105}, 257001  (2010).
\bibitem{mertelj} L. Stojchevska, P. Kusar, T. Mertelj, V. V. Kabanov, X. Lin,
G. H. Cao, Z. A. Xu, and D. Mihailovic, Phys. Rev. B {\bf 82}, 012505 (2010).
\bibitem{rettig} L. Rettig, R. Cortes, S. Thirupathaiah, P. Gegenwart, H.S. Jeevan,
M. Wolf, J. Fink, and U. Bovensiepen, Phys. Rev. Lett. {\bf 108}, 097002  (2012).
\bibitem{gianetti} S. Dal Conte, C. Giannetti, G. Coslovich, F. Cilento, D. Bossini, T. Abebaw,
F. Banfi, G. Ferrini, H. Eisaki, M. Greven, A. Damascelli, D. van der Marel, F Parmigiani,
Science {\bf 335}, 1600 (2012).
\bibitem{kaganov} M. I. Kaganov, I. M. Lifshits, and L. B. Tanatarov,
Zh. Eksp. Teor. Fiz., {\bf 31}, 232, (1956)  [Sov.Phys. JETP {\bf 4}, 173 (1957)].
\bibitem{anisimov} S.I. Anisimov, B.L. Kapeliovich, and T.L. Perel'man, Zh. Eksp. Teor. Fiz. {\bf 66},776 (1974).
[Sov. Phys. JETP, {\bf 39}, 375 (1974)]
\bibitem{allen} P. B. Allen, Phys. Rev. Lett. {\bf 59}, 1460 (1987).


\bibitem{eliashberg2} G. M. Eliashberg, \textit{Sov. Phys. JETP} \textbf{11}, 696 (1960) [Zh. Eksp. Teor. Fiz. \textbf{38}, 966 (1960)].

\bibitem{eliashberg4} G. M. Eliashberg, \textit{Sov. Phys. JETP} \textbf{12}, 1000 (1960) [Zh. Eksp. Teor. Fiz. \textbf{38}, 966 (1960)].

\bibitem{kabanovAlex} V.V. Kabanov, A.S. Alexandrov, Phys. Rev. B {\bf 78}, 174514 (2008).
\bibitem{Fann} W.S. Fann, R. Storz, H.W.K. Tom, and J. Bokor, Phys. Rev. B, {\bf 46}, 13592  (1992).
\bibitem{lisowski} M. Lisowski, P.A. Loukakos, U. Bovensiepen, J. Stahler, C. Gahl, and M. Wolf,
Appl. Phys. A {\bf 78}, 165 (2004).
\bibitem{perfetti} L. Perfetti, P.A. Loukakos, M. Lisowski, U. Bovensiepen, H. Eisaki and M. Wolf,
Phys. Rev. Lett. {\bf 99}, 197001 (2007).
\bibitem{Hufner} F. Reinert, S. H$\mathrm{\ddot{u}}$efner, in Very High Resolution Photoelectron Spectroscopy, edited by
S. H$\mathrm{\ddot{u}}$efner (Springer, Berlin Heidelberg, 2007.
\bibitem{jure} J. Demsar, R. D. Averitt,  A. J. Taylor,  V.V. Kabanov,  W. N. Kang,  H. J. Kim,
E.M. Choi,  and S. I. Lee, Phys. Rev. Lett. {\bf 91}, 267002 (2003).
\bibitem{primoz} P. Kusar, V.V. Kabanov, S. Sugai, J. Demsar, T. Mertelj, and D. Mihailovic,
Phys. Rev. Lett. {\bf 101}, 227001 (2008).
\bibitem{devereaux} M. Sentef, A. F. Kemper, B. Moritz, J. K. Freericks,
Z.-X. Shen, and T. P. Devereaux, Phys. Rev. X, {\bf 3}, 041033 (2013); J. A. Sobota, S.-L. Yang, D. Leuenberger,
A. F. Kemper, J. G. Analytis, I. R. Fisher, P. S. Kirchmann, T. P. Devereaux, and Z.-X. Shen, arXiv:1401.3078.
\bibitem{stat_fiz2}E.M. Lifshitz, L.P. Pitaevskii, Statistical Physics, Part 2 (Pergamon Press, Oxford 1980)
\bibitem{SmithJensen}H. Smith, H.H. Jensen Transport phenomena, Clarendon Press, Oxford, 1989.
\bibitem{fiz_kin} E.M. Lifshitz, L.P. Pitaevskii, Physical Kinetcs (Pergamon Press, Oxford 1981)
\bibitem{poor1} D. Belitz, Phys. Rev. B, {\bf 36}, 2513 (1987);
D. Belitz and M. N. Wybourne, Phys. Rev. B, {\bf 51}, 689 (1995);
B. I. Belevtsev, Yu. F. Komnik, and E. Yu. Beliayev, Phys. Rev.
B, {\bf 58}, 8079 (1998).
\bibitem{gusevWright} V.E. Gusev, O.B. Wright Phys. Rev. B, {\bf 57}, 2878 (1998).
\bibitem{elesin} V. F. Elesin, Yu. V. Kopaev, Sov. Phys. Uspekhi, {\bf 24}, 116 (1981).
\bibitem{tas} G. Tas, H.J. Maris Phys. Rev. B, {\bf 49}, 15046 (1994).
\bibitem{beyer} M. Beyer, D. Stadter, M. Beck, H. Schafer, V. V. Kabanov, G. Logvenov, I. Bozovic, G. Koren, and J. Demsar, Phys. Rev. B, {\bf 83}, 214515 (2011).
\end{thebibliography}
\end{document}